\documentclass{article}
\usepackage{waspaa21,amsmath,graphicx,url,times, enumitem}
\usepackage{color}
\usepackage{booktabs}
\usepackage{gensymb}
\usepackage{multirow}
\usepackage{array}
\usepackage{cite}
\newcolumntype{P}[1]{>{\centering\arraybackslash}p{#1}}
\def \am [#1]{\textcolor{red}{AM: #1}}
\def \rev [#1]{\textcolor{blue}{r: #1}}
\def \dk [#1]{\textcolor{green}{DK: #1}}
\title{Joint Direction and Proximity Classification of Overlapping Sound Events from Binaural Audio}

\name{Daniel Aleksander Krause$^{1}$, Archontis Politis$^{2}$, Annamaria Mesaros$^{3}$ \thanks{ This paper was partially supported by Academy of Finland grant 332063 "Teaching machines to listen".
}}
\address{Faculty of Information Technology and Communication Sciences, Tampere University\\
Korkeakoulunkatu 7, 33720, Tampere, Finland,\\
$^1$ daniel.krause@tuni.fi,           
         $^2$ archontis.politis@tuni.fi,
         $^3$ annamaria.mesaros@tuni.fi}

\begin{document}

\ninept
\maketitle

\begin{sloppy}

\begin{abstract}
Sound source proximity and distance estimation are of great interest in many practical applications, since they provide significant information for acoustic scene analysis. As both tasks share complementary qualities, ensuring efficient interaction between these two is crucial for a complete picture of an aural environment. In this paper, we aim to investigate several ways of performing joint proximity and direction estimation from binaural recordings, both defined as coarse classification problems based on Deep Neural Networks (DNNs). Considering the limitations of binaural audio, we propose two methods of splitting the sphere into angular areas in order to obtain a set of directional classes. For each method we study different model types to acquire information about the direction-of-arrival (DoA). Finally, we propose various ways of combining the proximity and direction estimation problems into a joint task providing temporal information about the onsets and offsets of the appearing sources. Experiments are performed for a synthetic reverberant binaural dataset consisting of up to two overlapping sound events.
\end{abstract}


\section{Introduction}
\label{sec:intro}
Acoustic environments consist of numerous types of sounds that are produced by sources distributed across space. Research in acoustic scene analysis has led to the formation of several inter-related audio tasks, including sound event detection and acoustic scene classification \cite{DCASE2016}. Localization and tracking of acoustic sources \cite{brandstein2001microphone} are some of the oldest and most researched tasks, and constitute an important part of many practical applications like surveillance systems \cite{Kotus_Jozef_Application_2011}, audio-driven robotics \cite{HornsteinHRTF}, teleconferencing \cite{AokiTeleconf}, or speech recognition and enhancement \cite{wolfel2009speech,virtanen2012speech}. Recent research has led to an increased focus on merged tasks utilizing spatial recordings, such as sound event detection and localization (SELD) \cite{Adavanne_2019}.
% am: cite here our journal?%
%Although source distance and direction-of-arrival (DOA) estimation are strictly interrelated, the former has been investigated to a much lesser extent \cite{BrendelDistance}. \dk[Hope it's more clear now:]{} \rev[Therefore, joint performance of these tasks should be a relevant part of the current research on acoustic scene analysis.]
% \rev[Therefore, joint performance of these tasks should be a natural part of a complex acoustic scene analysis system.]
Although source distance and direction-of-arrival are both estimated from spatial information captured between the microphones, the former has been investigated to a much lesser extent \cite{BrendelDistance}, due to its estimation being crippled for distances that are a few times larger than the array size. In the case of binaural localization, this region extends to no more than about a meter from the listener, and understanding whether a source is proximal to the listener or not is still of interest in many applications.

%Most common approaches that tackle both problems rely on microphone arrays, which allow for precise multi-channel audio recordings \cite{Gburrek2020, Kotus2013MultipleSS, perotin2019regression}. However, this approach is limited by certain recording conditions and portability of the gear, which might not be suitable for many applications. A common and convenient alternative are binaural cues, which require only a pair of microphones, whilst utilizing spatial information provided by the Head Related Transfer Function (HRTF) \cite{Xie2013HeadRelatedTF}.
The majority of localization methods rely on multi-microphone arrays with more than two microphones, employing linear, circular, or spherical arrangements, depending on the application. However, two channel binaural recordings are an important format for acoustic scene analysis, mainly because they are based on the same spatial cues as the auditory system, they match the audio perspective of the human or human-like recorder, and they are a popular option in anthropomorphic robotic processing \cite{murray2004robotics,trowitzsch2017robust}.

Early works on binaural localization and distance estimation have focused on statistical methods using binaural cues \cite{Keyrouz2006,Rodemann2010ASO}. More recent studies utilize deep neural networks to overcome the limitations of model-based approaches and improve robustness to interfering noise and reverberant conditions \cite{May2012,Trowitzsch2020}. However, most studies investigating joint localization and distance estimation have focused on the azimuth plane only \cite{Yiwere2017DistanceEA,Ding2020JointEO}, since inter-channel binaural cues contain inherent ambiguities with elevation due to the cone-of-confusion effect \cite{blauert1997spatial}. Some studies analyze performance on elevation classification separately from azimuth \cite{Thuillier2018}, whereas other studies include joint azimuth and elevation localization, but restricted to certain ranges, e.g., the frontal area \cite{Pang2019}. Analogously, most work focused on distance estimation is restricted to sources at fixed positions \cite{Zohourian2020} or varying in the azimuth plane only \cite{Zohourian2019}.

In this paper, we propose methods for obtaining joint proximity and direction information from binaural audio. We treat both problems as coarse classification tasks, in which proximity is described in a binary way, i.e., as near or far. For direction, two ways of splitting the sphere into angular areas are proposed, in order to obtain a set of rough classes - left, right, front, back, top and bottom. In the first part of our study, we investigate several ways of performing direction classification utilizing a single-label and a multi-label approach. In the second part, different techniques of combining the proximity and direction classifications tasks are investigated. Compared to existing studies on these topics, we extend our research to a scenario with temporally overlapping sound events spread across the whole sphere, and diverse reverberation conditions. Hence, models are trained and evaluated at a frame level to provide information about the onsets and offsets of the  sound sources. To the authors' knowledge, this scenario has not been studied so far for binaural audio.

\section{Method}
\label{sec:method}

\subsection{Model and features}

\renewcommand{\arraystretch}{0.7}
\begin{figure}[!t]
\centerline{\includegraphics[width=60mm,scale=0.5]{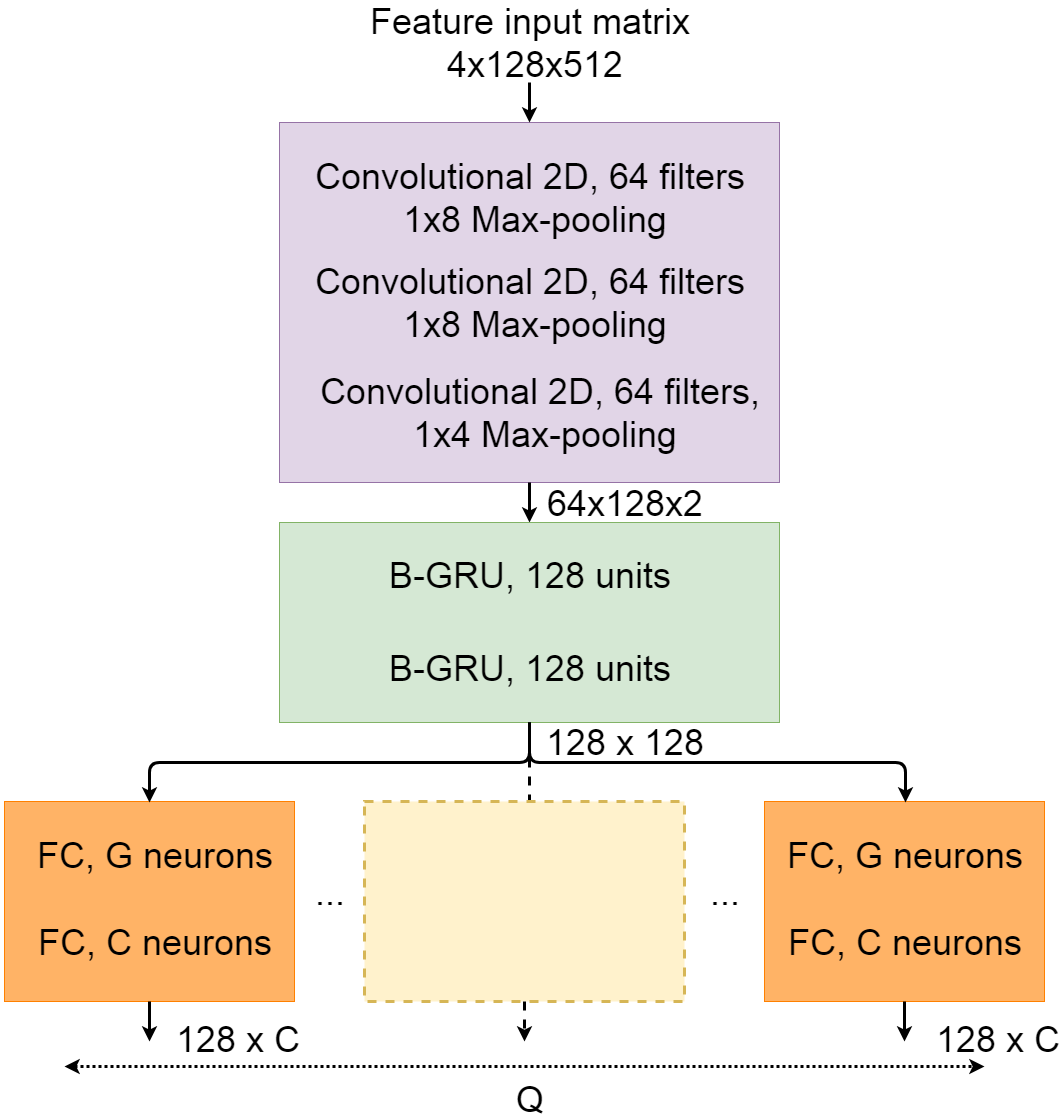}}
\caption{Baseline architecture used across all experiments.}
\label{fig_model}
\end{figure}
\renewcommand{\arraystretch}{1.0}
To enable concurrent spatial information processing and temporal detection of sound sources, we utilize a convolutional recurrent neural network (CRNN) architecture shown in~Fig.~ \ref{fig_model}. This kind of model architecture has been proven to be effective in similar audio tasks \cite{Adavanne_2019, krausefeat}. The first block of the model consists of three convolutional layers with 3x3 kernels. After each convolutional part, ReLU activation outputs are followed by batch normalization and MaxPooling across frequency. The convolutional block is followed by two bidirectional gated recurrent units (B-GRU), each scaled by a hyperbolic tangent function. The bottom part of the model consists of $Q$ fully connected (FC) branches. Each branch contains a $G$-dimensional linear layer, followed by an output layer with $C$ neurons and a sigmoid activation function.

% \begin{figure}[!t]
% \centerline{\includegraphics[width=60mm,scale=0.5]{Model arch.png}}
% \caption{Architecture baseline for the models used across all experiments.}
% \label{fig_model}
% \end{figure}

% FEATURES 
  Models are trained using a set of three different feature representations. Firstly, we compute the complex spectrogram using a short-time Fourier transform with a Hamming window of 40$ms$ length and 50\% overlap. Next, to represent the phase shift between binaural channels we compute the sine and cosine values of interaural phase differences (\textbf{sin\&cos}). These features have been shown to outperform direct use of phase differences in both speech separation \cite{sincosfeat} and sound event localization \cite{krausefeat}, as they avoid phase wrapping and provide a smoother representation of highly varying phase values.  Furthermore, we utilize interaural level differences (\textbf{ILDs}) which, considering head shadowing effects, become significant at frequencies above 1.5kHz and constitute one of the two major cues in binaural localization \cite{blauert1997spatial}. The second major binaural localization cue, time-difference-of-arrival, which is mostly dominant at frequencies below 1kHz, is represented in the features by the phase differences to which it relates.
%   \begin{gather}
%     \mathrm{SI}[n,k]=\sin{(\mathrm{IPD}[n,k])},
%     \\
%     \mathrm{CI}[n,k]=\cos{(\mathrm{IPD}[n,k])}.
% \end{gather}
% The inter-channel phase differences (IPDs) for the left and right channel (denoted as $l$ and $r$) are defined as:
% \begin{gather}
%     \mathrm{IPD}[n,k]=\arg(X_l[n,k])-\arg(X_r[n,k]).
% \end{gather}
% where $X_i[n,k]$ denotes the complex spectrum value, $\arg(X_i[n,k])$ represents the phase spectrogram, and $n$ and $k$ denote the time and frequency indices, respectively. Sin\&cos features have been shown to outperform standard phase difference in both speech separation \cite{sincosfeat} and sound event localization \cite{krausefeat}. Furthermore, we utilize inter-channel level differences (ILDs) which become significant at frequencies above 1.5kHz \cite{BlaurtSpatialHearing}, defined as:
% \begin{gather}
%     \mathrm{ILD}[n,k]=\frac{|X_l[n,k]|}{|X_r[n,k]|}.
% \end{gather}
Interaural level differences can flatten source-related magnitude patterns, which may contain important information for temporal detection of events. Therefore, a single magnitude spectrogram is added to offset this deficiency, making for 4 feature channels in total. The modelled sequence consists of 128 frames, which results in a 4x128x512 input matrix.
DNNs are trained using the Adam optimizer \cite{ruder2017overview} and the Keras library \cite{chollet2015keras}, with binary-cross entropy used as the loss function. A single output is considered to be positive when its value is greater than or equal to $0.5$.

\subsection{Data}
Experiments are performed using a synthetic dataset. In order to ensure a diverse representation of sound sources, data is created using isolated sound events derived from several datasets: NIGENS \cite{trowitzsch_ivo_2019_2535878}, DESED\cite{turpault:hal-02160855} and TUT Rare Sound Events 2017 \cite{DCASE2017challenge}, containing 18 total sound classes, namely: alarm, baby, blender, cat, crash, dishes, dog, engine, fire, footsteps, glassbreak, gunshot, knock, phone, piano, scream, speech, water. The data is split into two subsets, one of which contains up to two overlapping sound events, whereas the other one consists of single sources only. Each subset contains 400 audio files, divided into 4 equal splits for fold-wise cross-validation. 

In order to provide a possibly large and general representation of localization scenarios, we simulate a random shoe-box room for each file separately. We randomize both the reverberation time (RT) and physical dimensions of the room, as well as the receiver position. For each sound source, a separate room impulse response (RIR) is synthesized by selecting a random location inside the room. Randomization parameters are summarized in Table \ref{tab:randomtab}. Binaural RIRs are simulated using the image source method as implemented in \cite{Allen1976ImageMF}, where individual image sources are convolved with head-related impulse responses (HRIRs) from the respective source direction. We employ an available HRIR set that includes near-field HRIRs \cite{qu2009distance}, measured at every 10cm radiuses away from the head, from 20cm up to 1.5 m. In that way it is possible to include all the relevant near-field distance cues of proximal sources. If the source is closer than 1.5 m the respective distance-dependent HRIR is convolved with the direct sound part, while the rest of the image sources are convolved with the far-field 1.5 m HRIR set. Each source and receiver is forced to be at least 1 m apart from the walls, ceiling, and floor. The recordings are finally synthesized by convolving randomly picked sound events with the corresponding RIRs, with their temporal onset placed randomly in the recording. Each audio file is 15 seconds long with frequency sampling of 24 kHz and 16 bit resolution. The dataset is available at Zenodo \footnote{10.5281/zenodo.5118587}.
\renewcommand{\arraystretch}{0.7}
\begin{table}[!t]
  \centering
  \caption{Randomization of parameters for data generation.}
    \begin{tabular}{cc}
    \toprule
    Parameter   & Random range  \\
    \midrule
    \midrule
    Room width and length & [6.0 10.0] m  \\
    Room height & [2.5 6.0] m  \\
    RT & [0.3 0.9] s \\
    Source distance (near)  & [0.4 2.0] m \\
    Source distance (far)  & [3.0 8.0] m \\
    \bottomrule
    \end{tabular}%
  \label{tab:randomtab}%
\end{table}%
\begin{table}[!t]
  \centering
  \caption{Definition of the equal and unequal sphere division types.}
    \begin{tabular}{P{0.15\linewidth}P{0.1\linewidth}P{0.25\linewidth}P{0.25\linewidth}}
    \toprule 
    Division & Direction & Azimuth [$^{\circ}$] & Elevation [$^{\circ}$]  \\
    \midrule
    \midrule
    \multirow{6}{*}{Unequal} & Front & [-45, 45] & [-35, 35] \\ 
    & Back & [135, -135] & [-35, 35] \\
    & Left & [45, 135] & [-35, 35] \\ 
    & Right & [-45, -135] & [-35, 35] \\
    & Top & [-180, 180] & [35, 90] \\
    & Bottom & [-180, 180] & [-90, -35] \\
    \midrule\midrule
        \multirow{6}{*}{Equal} & Front & [-90, 90] & [-90, 90] \\ 
    & Back & [90, -90] & [-90, 90] \\
    & Left & [0, 180] & [-90, 90] \\ 
    & Right & [-180, 0] & [-90, 90] \\
    & Top & [-180, 180] & [0, 90] \\
    & Bottom & [-180, 180] & [-90, 0] \\
    \bottomrule
    \end{tabular}%
  \label{tab:spheredivision}%
\end{table}%
\renewcommand{\arraystretch}{1.0}

\section{Experimental setup}
\label{sec:problem}
We investigate several methods to obtain joint information about the proximity and direction of sound sources from binaural audio. We treat both problems as coarse classification tasks with predefined classes. Proximity is expressed by two classes: near and far. The \textbf{near} class is defined for sources appearing in the proximal region around the head in the range of [0.4, 2.0] m. After some initial experiments, a buffer zone of 1m is added to avoid issues with sources appearing at the border of two classes. Therefore, the \textbf{far} class covers the distance from 3m upwards.

The direction task is defined for a coarse set of classes, corresponding to six basic ranges of DoAs - \textbf{left}, \textbf{right}, \textbf{top}, \textbf{bottom}, \textbf{front} and \textbf{back}. We propose two ways of defining those, as summarized in Table \ref{tab:spheredivision}:
\begin{itemize}[noitemsep]
    \item \textbf{Unequal:} this division type separates the sphere into six disjunctive cones, which are supposed to roughly reflect the intuitive human perception of each direction.
    \item \textbf{Equal:} %contrary to the former one, 
    this method divides the sphere into three sets of hemispheres defined around each of the Cartesian planes, instead of creating exclusive angular areas. Every source can be essentially described in each plane by one of two disjunctive classes - left or right, front or back and top or bottom, allowing for multi-label classification.
\end{itemize}

In the experiments, we investigate joint modelling of proximity and direction classification. Among these two tasks, classifying the direction appears to be more difficult due to its more complex nature and the spatial limitations of binaural recordings. Therefore, we split our experiments into two consecutive stages:
\begin{enumerate}[nolistsep]
    \item \textbf{Direction classification:} In this stage, we separately investigate  direction classification. We compare the performance of several ways to utilize both sphere divisions and perform the task using a single-label and multi-label approach.
    \item \textbf{Joint proximity and direction classification:} Here we combine the solutions analyzed in the previous steps with proximity classification. The joint task is performed as a single task  for the unequal sphere split and using a multi-task approach for both division variations.
\end{enumerate}
% \begin{itemize}
%     \item \textbf{Direction classification:} In this stage, we separately investigate  direction classification. We compare the performance of several ways to perform this as a multilabel task operating on the equal sphere division - by utilizing separate models for each plane and a single model with separate branches. We also show results obtained for a single label approach for mutually exclusive classes for both types of sphere divisions.
%     \item \textbf{Joint proximity and direction classification:} Here we combine the solutions analyzed in the previous steps with proximity classification. The joint task is performed as a single task  for the unequal sphere split and using a multi-task approach for both division variations.
% \end{itemize}

\subsection{Stage 1: Direction classification}
For both types of sphere divisions we propose several ways to perform direction classification:
\begin{itemize}[noitemsep]
    \item \textbf{UneqOne:} using the unequal split, we train a single model to perform classification for all 6 directions classes using a single output layer. In this case $Q=1, G=128, C=6$.
    \item \textbf{EqSepMod:} for the equal sphere division, we train three distinct models to perform direction detection in each hemisphere separately, resulting in a multi-label task as a whole. Hence, an exclusive binary classifier determines whether a sound source appears in the left or right area, while the other ones analyse the front/back and top/bottom planes. Since each output represents only two classes, we set lower values for the parameters: $Q=1, G=64, C=2$.
    \item \textbf{EqSepBran:} in this method, we use the same multi-label approach as in EqSepMod, however instead of training three models, we utilize a single DNN with separate output branches, each corresponding to a different hemisphere division. Hence, $Q=3, G=64, C=2$.
    % \item \textbf{EqOne:} to provide a reference method for the multi-label approach, we propose a neural network which performs single-label classification for combined classes, derived from the equal  division. Essentially, this separates the sphere into equal eighths, each being a combination of the hemispheric classes, e.g. top-front-left. The parameters for the output block are $Q=1, G=128, C=8$.
    \item \textbf{EqOne:} we use the equal division to separate the sphere into disjunctive eighths, each being a combination of the hemispheric classes, e.g. top-front-left. This concept is closer to UneqOne, since it is defined as single-label classification of exclusive directions that serve as a reference for results obtained with the multi-label approach. We perform classification for each of the 8 disjunctive classes using a single-label approach, hence $Q=1, G=128, C=8$.
\end{itemize}

We note that EqSepMod and EqSepBran give ambiguous information about the location of the sound source. To enable direct comparison of all methods, we combine the three independent outputs into 8 disjunctive classes as defined for EqOne. To achieve this, we use a frame-level probability product:
\begin{gather}
    \mathrm{P}_c[n]=\sqrt[3]{P_{lr}[n]*P_{fb}[n]*P_{tb}[n]} \geq 0.5.
\end{gather}
$P_c[n]$ denotes the output values for the $n$-th frame and $c$-th single-label class defined on the equal sphere division. Each value of $c$ corresponds to a unique combination of $lr, fb$ and $tb$, which are associated with probabilities obtained in the left/right, front/back and top/bottom planes respectively.

\subsection{Stage 2: Joint task}
\label{subsec:joint}

In the second phase of our experiments, we focus on joint modelling of proximity and direction classification. Here, we introduce the following methods of performing this task:
\begin{itemize}[noitemsep]
    \item \textbf{UneqSingle:} for the unequal sphere division, we combine all proximity and direction categories into a single-task set of 12 classes (e.g., near-left). Hence, $Q=1, G=128, C=12$.
    \item \textbf{UneqMulti:} alternatively, we propose a multi-task approach by utilizing a single fully connected layer with outputs corresponding to each task separately. Since there are 2 proximity classes and 6 direction classes, this results in the following parameters: $Q=1, G=128, C=8$.
    \item \textbf{EqSepMod-J:} this method combines the separate models trained for EqSepMod in the previous stage and utilizes an analogous fourth DNN to perform proximity classification.
    \item \textbf{EqSepBran-J:} similarly to the previous one, we use a model derived from EqSepBran with another output branch associated with proximity classes. Thus, $Q=4, G=64, C=2$.
    \item \textbf{EqOne-J:} analogously to UneqMulti, we combine proximity and direction classes into a multi-task output branch, utilizing the equal sphere division. $Q=1, G=128, C=10$.
    
\end{itemize}

\section{Results and discussion}
\label{sec:results}
\subsection{Direction classification}

Table \ref{tab:1source} and \ref{tab:2sources} show results obtained for data with one source and two concurrent sources. Model performance is measured using the one second segment-based $F_{1}$ score averaged over all cross-validations folds.

As can be observed for the results obtained on the unequal sphere split (UneqOne), there is a significant imbalance between the model's performance on different classes. With an overall $F_{1}$ measure of 55.30$\%$ (Table \ref{tab:2sources}), the model scores at 71.77$\%$ for the left/right classes, whereas for front/back the performance decreases to 48.35$\%$. Overall performance is strongly affected by the top/bottom directions, for which the accuracy scored at 26.40$\%$. This contrast is inherently related to the binaural format qualities. Largest differences between channels appear in the left/right plane, which allows easier localization. For the front/back plane the distinction becomes less pronounced due to diminishing interaural differences. The top and bottom directions are affected most by binaural limitations, with elevation localization cues known to be weaker than lateralization cues \cite{blauert1997spatial}. 

% As can be observed for the results obtained on the unequal sphere split (UneqOne), there is a significant imbalance between the model's performance on different classes. With an overall $F_{1}$ measure of 55.30$\%$ (\ref{tab:2sources}), the model scores at 71.77$\%$ for the left/right classes, whereas for front/back the performance decreases to 48.35$\%$. Overall performance is strongly affected by the top/bottom directions, for which the model achieved the worst accuracy - 26.40$\%$. These differences are inherently related to the characteristics of the binaural format. Largest differences between channels appear in the left/right plane, making it the easiest to localize sound sources. For the front/back plane the distinction becomes less pronounced due to the cone of confusion effect. The top and bottom directions are affected the most by binaural limitations, since elevation resolution of HRTFs is known to be significantly worse in comparison with the azimuth plane. 

Notable improvements can be obtained by using a multi-label approach and the equal sphere division. Utilizing separate models (EqSepMod) for each plane results in a high score for the left/right directions (88.29$\% F_{1}$). The front/back and top/bottom models achieve 74.76$\%$ and 74.24$\%$, outperforming the UneqOne model by 26.41 p.p. and 47.84 p.p. respectively. We note that in this method each class includes a whole hemisphere of cues, contrary to the unequal sphere division in which all directions are defined by narrower cones. Interestingly, a single model with separate branches (EqSepBran) shows just a very slight decrease in performance of about 1 p.p. in each plane, indicating that very similar results can be achieved with a joint DNN. EqSepBran and EqSepMod obtain overall multi-label scores of 78.05$\%$ and 79.11$\%$. However, combining the results for each plane into a single-label grid using (1) shows a respective drop to 46.79$\%$ and 47.90$\%$. Despite efficient direction classification in each dimension, different onsets and offsets might create a challenge with temporal matching of independently obtained output sequences. This approach might therefore benefit from a more intelligent approach than a frame-level probabilistic product, which achieves slightly worse results than UneqOne. Training a model directly with a single-label approach (EqOne) results in 34.11$\% F_{1}$ measure, which is the worst overall score. Compared with UneqOne, the performance might be affected by a different sector orientation and a finer division of the sphere. A difference of over 13 p.p. and high accuracy for the front/back and top/bottom classes show the potential of multi-label training for future research. 

Results obtained for both data subsets show coherent relations between methods for the single- and two-source case. No significant differences can be observed for the left and right directions, however we note a significant drop in the front/back and top/bottom planes for a single source scenario. We link it to a higher number of false positives caused by switching between opposite classes due to the cone of confusion. This effect might be partly masked in a double source scenario by the appearance of two sound events located on opposing sides of the head.
% the lack of a trackwise model output and an appearance of concurrent sound events.

\renewcommand{\arraystretch}{0.7}
\begin{table}[!t]
  \centering
  \caption{$F_{1}$ score [\%] obtained for a single source scenario.}
    \begin{tabular}{P{0.2\linewidth}P{0.15\linewidth}P{0.10\linewidth}P{0.15\linewidth}P{0.15\linewidth}}
    \toprule
    & UneqOne & EqOne & EqSepBran & EqSepMod  \\
    \midrule
    \midrule
    Left/Right & 71.39 & - & 88.25 &  \textbf{89.82} \\
    Front/Back & 39.91 & - & 60.54 & \textbf{65.75} \\
    Top/Bottom & 28.66 & - & 60.53 & \textbf{63.75}  \\\cmidrule{2-5}
    Single-label & \textbf{51.98} & 30.44 & 29.66 & 34.37\\\cmidrule{2-5}
    
    Multi-label & - & - & 68.76 & \textbf{73.11} \\
    \bottomrule
    \end{tabular}%
  \label{tab:1source}%
\end{table}%

\begin{table}[!t]
  \centering
  \caption{$F_{1}$ score [\%] obtained for two concurrent sources.}
     \begin{tabular}{P{0.2\linewidth}P{0.15\linewidth}P{0.10\linewidth}P{0.15\linewidth}P{0.15\linewidth}}
    
    \toprule
   & UneqOne & EqOne & EqSepBran & EqSepMod  \\
    \midrule
    \midrule
    Left/Right & 71.77 & - & 87.45 &  \textbf{88.29} \\
    Front/Back & 48.35 & - & 73.88 & \textbf{74.76} \\
    Top/Bottom & 26.40 & - & 73.10 & \textbf{74.24}  \\\cmidrule{2-5}
    Single-label & \textbf{55.30} & 34.11 & 46.79 & 47.90 \\\cmidrule{2-5}
    Multi-label & - & - & 78.05 & \textbf{79.11} \\
    \bottomrule
    \end{tabular}%
  \label{tab:2sources}%
\end{table}%
\renewcommand{\arraystretch}{1.0}
\subsection{Joint proximity and direction classification}
Results obtained for the joint task of proximity and direction classification are summarized in Table \ref{tab:joint}. Combinations marked with a "-J' ending refer to models evaluated in the previous stage with an additional branch performing the proximity task. UneqMulti and UneqSingle are models described in section \ref{subsec:joint}.
% UneqMulti represents a DNN with a shared output layer, but separate cells for each task, trained in a multi-task manner. UneqSingle refers to a single-task approach, in which the output consists of all direction and proximity class combinations.

\renewcommand{\arraystretch}{0.7}
\begin{table}[!t]
  \centering
  \caption{$F_{1}$ score [\%] obtained for the joint task. Direction performance is described for a double source single-label scenario. The last column shows separate direction classification for reference.}
     \begin{tabular}{P{0.2\linewidth}P{0.15\linewidth}P{0.15\linewidth}P{0.209\linewidth}}
    \toprule
    Model & Proximity & Direction & Sep. direction  \\
    \midrule
    \midrule
    % \multirow{6}{*}{One source} & UneqSingle & \multicolumn{2}{c}{43.01} \\ & UneqSepMod-J & \textbf{86.57} & \textbf{51.98} \\ & UneqMulti & 83.49 & 48.94 \\ \cmidrule{2-4} & EqOne-J & 82.95 & 21.77 \\ & EqSepBran-J & 83.12 & 29.81 \\ & EqSepMod-J & \textbf{86.57} & \textbf{34.37} \\ \midrule\midrule
    UneqSingle & \multicolumn{2}{c}{50.63} & \multirow{2}{*}{55.03} \\ UneqMulti & 82.54 & 51.03 & \\ 
    \cmidrule{2-4} EqOne-J & 83.45 & 30.31 & 34.11 \\ EqSepBran-J & \textbf{83.92} & \textbf{45.39} & 46.79\\
    EqSepMod-J & \textbf{83.97} & \textbf{47.90} & 47.90\\
    \bottomrule
    \end{tabular}%
  \label{tab:joint}%
\end{table}%
\renewcommand{\arraystretch}{1.0}
As a reference for the tested network variations, we use results derived from separate DNNs. A model trained for proximity classification alone achieves a score of \textbf{83.97$\%$}, which shows that coarse proximity classification in conjunction with temporal detection of sound sources can be learnt very efficiently using binaural audio. Considering the connection of proximity information with other audio tasks, our results set a space for wider research in the future. 

A decline in performance of both tasks when combined into a joint model is well expected, however the differences in our experiments are not major. Specifically, the proximity task shows a maximal drop of only 1.34 p.p. when combined using the multi-task method (UneqMulti). The lowest performance difference for both tasks can be observed using separate multi-label branches (EqSepBran-J). Compared with the baseline models, $F_{1}$ score decreases by 0.05 p.p. and 1.4 p.p. for proximity and direction, respectively. These results are comparable with the separate model approach (EqSepMod-J). We notice a more significant decrease of direction classification for a single-label approach, for both equal (EqOne-J) and unequal sphere division (UneqMulti). This might suggest that multi-label learning is less prone to be affected by joint tasks. Finally, utilizing a single-task output for the equal sphere split (UneqSingle) results in 50.63$\% F_{1}$ score, which is mostly affected by poor direction classification. Since this approach disables direct separation of information about proximity and direction, its low efficiency makes it the least practical solution in this setting.

\section{Conclusions}
\label{sec:conclusions}

In this paper, we investigate a scenario in which joint proximity and direction classification is performed in conjunction with temporal onset and offset detection of overlapping sound events. We propose two ways of splitting the whole sphere into coarse sets of direction classes, for which we propose a single-label and multi-label approach to perform classification. Finally, we present several ways to combine the proximity and direction classification tasks to perform them in a joint manner. 

The main challenge when tackling the direction classification problem appears in the top/bottom and front/back planes, that are most affected by the cone of confusion. Our experiments show that with the proposed multi-label approach and a hemisphere division of classes the $F_{1}$ score for these directions can be increased significantly. In this case, combining independent output sequences into disjunctive areas can become another problem, which is a subject for future research. Finally, we show that joint proximity and direction classification can be achieved without significant losses in performance for both problems, with the multi-task approach standing out as the most efficient one.

% funding goes to thanks
% \section{ACKNOWLEDGMENT}
% \label{sec:ack}
% This paper has received funding from Academy of Finland grant 332063 "Teaching machines to listen".

% -------------------------------------------------------------------------
% Either list references using the bibliography style file IEEEtran.bst
\bibliographystyle{IEEEtran}
\bibliography{2021_WASPAA_PaperTemplate_Latex}
%
% or list them by yourself
% \begin{thebibliography}{9}
% 
% \bibitem{waspaa21web}
%   \url{http://www.waspaa.com}.
%
% \bibitem{IEEEPDFSpec}
%   {PDF} specification for {IEEE} {X}plore$^{\textregistered}$,
%   \url{http://www.ieee.org/portal/cms_docs/pubs/confstandards/pdfs/IEEE-PDF-SpecV401.pdf}.
%
% \bibitem{PDFOpenSourceTools}
%   Creating high resolution {PDF} files for book production with 
%   open source tools, 
%   \url{http://www.grassbook.org/neteler/highres_pdf.html}.
%
% \bibitem{eWilliams1999}
% E. Williams, \emph{Fourier Acoustics: Sound Radiation and Nearfield Acoustic
%   Holography}. London, UK: Academic Press, 1999.
% 
% \bibitem{ieeecopyright}
%   \url{http://www.ieee.org/web/publications/rights/copyrightmain.html}.
%
% \bibitem{cJones2003}
% C. Jones, A. Smith, and E. Roberts, ``A sample paper in conference
%   proceedings,'' in \emph{Proc. IEEE ICASSP}, vol. II, 2003, pp. 803--806.
% 
% \bibitem{aSmith2000}
% A. Smith, C. Jones, and E. Roberts, ``A sample paper in journals,'' 
%   \emph{IEEE Trans. Signal Process.}, vol. 62, pp. 291--294, Jan. 2000.
% 
% \end{thebibliography}

\end{sloppy}
\end{document}